\documentstyle[sprocl,epsf]{article}

\begin{document}

\def\ii{\'{\char'20}}
\def\bea{\begin{eqnarray}}
\def\eea{\end{eqnarray}}
\newcommand{\beq}{\begin{equation}}
\newcommand{\eeq}{\end{equation}}

\title{Polchinski ERG Equation and 2D Scalar Field Theory \footnote{
Talk given by Rui Neves at the Workshop on the Exact Renormalization Group\\ 
\hbox{}\hspace{0.6cm}(Faro, September 10-12 1998).}}
\author{Yuri Kubyshin \footnote{On leave of absence from the Institute 
for Nuclear Physics, Moscow State University, \\
\hbox{}\hspace{0.6cm} 117899 Moscow, Russia.}, 
Rui Neves and Robertus Potting\\
\footnotesize{Department of Physics,
Universidade do Algarve, 8000 Faro, Portugal}}

\maketitle

\begin{abstract}
We investigate a $Z_2$-symmetric scalar field theory in two 
dimensions using the Polchinski exact renormalization group 
equation expanded to second order in the derivative expansion. 
We find preliminary evidence that the Polchinski equation 
is able to describe the non-perturbative infinite set of 
fixed points in the theory space, corresponding to the minimal 
unitary series of 2D conformal field theories. We compute 
the anomalous scaling dimension $\eta$ and the correlation 
length critical exponent $\nu$ showing that an accurate 
fit to conformal field theory selects particular 
regulating functions. 
\end{abstract}

\section{Introduction}
Since its very origins, the exact renormalization group (ERG) 
\cite{KW,SW,WH,JP} has proved to be a powerful tool for studies of 
non-perturbative effects in quantum field theory (see recent 
reviews in \cite{YK,TM1}).

A particularly interesting case is that of an effective scalar 
field theory in two dimensions. As it was first conjectured by 
Zamolodchikov \cite{AZ}, for a $Z_2$ symmetric theory there 
should exist an infinite set of non-perturbative fixed 
points corresponding to the unitary minimal series of $(p,p+1)$ 
conformal field theories, where $p=3,4,\ldots,\infty$ 
\cite{BPZ}. Morris \cite{TM} showed numerically that such 
points do exist.  
The calculation was performed with a reparametrization 
invariant version of the Legendre ERG equation \cite{TM2} 
expanded in powers of derivatives. It was also pointed out 
there that to the level of the local potential 
approximation only the continuum limits described by periodic 
solutions and corresponding to critical sine-Gordon models 
could be obtained. To find the expected set of fixed points the 
calculations had to be taken to the next order in the derivative 
expansion. This constituted a manifestation of the non-perturbative 
nature of the phenomena, and remarkably the Legendre ERG equation was 
powerful enough to locate and describe with good accuracy 
the expected set of 2D field theories.

In this work we study the same $Z_2$ symmetric scalar 
field theories in two dimensions but now with the 
Polchinski ERG equation \cite{JP}. We present preliminary results 
which complement the results obtained with the Legendre ERG 
equation. In Sect. 2 we follow the article by Ball 
et al. \cite{BHLM} to present the basic equations 
of the formalism. This will allow us to set up notation for 
Sect. 3 where we analyse the equations to second order in 
the derivative expansion. In Sect. 4 we discuss the results and 
present our conclusions.

\section{Polchinski equation and derivative expansion}

The Polchinski equation \cite{JP} for a scalar theory can be 
written as follows \cite{BT} 
\bea
 & & {\partial\over{\partial t}}\hat{S}={\int_{\hat{p}}}
 K'({\hat{p}^2})\left[{{\delta \hat{S}}\over{\delta{\hat{\varphi}_p}}}
 {{\delta \hat{S}}\over{\delta{\hat{\varphi}_{-p}}}}-
{{{\delta^2}\hat{S}}\over{\delta{\hat{\varphi}_p}
\delta{\hat{\varphi}_{-p}}}}\right]+dS+ \nonumber \\
 & & {\int_{\hat{p}}}\left[1-{d\over{2}}-{\eta(t)
 \over{2}}-2{\hat{p}^2}{{K'({\hat{p}^2})}\over{K({\hat{p}^2})}}\right]
 {\hat{\varphi}_p}{{\delta \hat{S}}\over{\delta{\hat{\varphi}_p}}} - 
 {\int_{\hat{p}}}{\hat{\varphi}_p}\hat{p}\cdot
{\partial'\over{\partial\hat{p}}}{{\delta \hat{S}}
\over{\delta{\hat{\varphi}_p}}}    \label{cv1}.
\eea
Here $\hat{S}$ is a general Wilsonian action which can be written 
in terms of dimensionless variables as follows
\bea
\hat{S}[\hat{\varphi};t] & = & \frac{1}{2} \int_{p} \hat{\phi}_{p} 
  p^{2} \left( K(\hat{p}^{2}) \right)^{-1} \hat{\phi}_{-p} + 
\hat{S}_{int}[\hat{\varphi};t],  \label{action1} \\
\hat{S}_{int}[\hat{\varphi};t] & = & \int dy \left[ v(\hat{\phi}(y),t) 
+ z(\hat{\phi}(y),t) \left( \frac{\partial \hat{\phi}}{\partial y^{\mu}}
\right)^{2} + \ldots \right]. 
\label{action2}  
\eea
In Eq.\ (\ref{cv1}) the partial derivative on $\hat{S}$ means it 
only acts on the explicit 
$t=\ln({\Lambda_0}/\Lambda)$ dependence 
of the couplings and the prime in the momentum derivative means 
it does not act on the delta function of the energy-momentum 
conservation, and ${\int_{\hat{p}}}\equiv\int{{{d^d}\hat{p}}
\over{{(2\pi)}^d}}$. $K(\hat{p}^2)$ is a (smooth) 
regulating function which damps the high energy modes satisfying 
the normalization condition $K(0)=1$.  
The renormalized field $\hat{\varphi}_p$ changes with 
scale according to
\begin{equation}
\Lambda \frac{\partial}{\partial \Lambda}\hat{\varphi}_{p}=
\left[ 1+{d\over{2}}-{1\over{2}}\eta(t) \right] \hat{\varphi}_{p},
\end{equation}
where $\eta(t)$ is the anomalous scaling dimension. 

To the second order in the derivative expansion we 
consider the two terms which are written explicitly in 
Eq.\ (\ref{action2}). Within this approximation the Polchinski ERG 
equation reduces to the following system \cite{BHLM} 
\bea
\dot{f} & = & f''+2Az'-2ff'+{\Delta^+}f+{\Delta^-}xf', 
\label{eq1} \\
\dot{z} & = & z''+B{{f'}^2}-4zf'-2z'f+{\Delta^-}xz'-
\eta z-\eta/2,   \label{eq2}
\eea
where ${\Delta^\pm}=1\pm d/2-\eta/2$, $f(x)=v'(x)$, $x\equiv\hat{\varphi}$ 
and the potentials $v(x,t)$ and $z(x,t)$ are defined in 
Eq.\ (\ref{action2}). The dots and primes denote the partial derivatives 
with respect to $t$ and $x$ respectively. The parameters $A$ and $B$ reflect 
the scheme dependence of the equations and are equal to
$A=({I_1}{K_0})/{I_0}$, $B={K_1}/{K_0^2}$.
Here $K_n$ and $I_n$, $n=0,1,\cdots$ parametrize the regulating  
function in Eq.\ (\ref{action1}) and are defined by 
\bea
{K_n} & = & {{(-1)}^{n+1}}{K^{(n+1)}}(0), \nonumber \\ 
{I_n} & = & -{\int_{\hat{p}}}{{({\hat{p}^2})}^n}K'
({\hat{p}^2})=-{\Omega_d}{\int_0^\infty}dz {z^{d/2-1-n}}K'(z),
  \nonumber
\eea
where $K^{(n)}$ stands for the $n$-th derivative of $K$ and 
${\Omega_d}=2/(\Gamma(d/2){{(4\pi)}^d})$. 

In the next section we search for fixed-point solutions, i.e. 
for functions $f(x)$ and $z(x)$ which are independent of $t$ 
and satisfy the system 
\bea
& & f''+2Az'-2ff'+{\Delta^+}f+{\Delta^-}xf'=0, \label{eqfp1} \\
& & z''+B{{f'}^2}-4zf'-2z'f+{\Delta^-}xz'-
\eta z-\eta/2=0.   \label{eqfp2}
\eea
We will choose the initial conditions (according to the terminology 
adopted in the literature on the ERG equations) set by the 
$Z_2$ symmetry: $f(0)=0$ and $z'(0)=0$ and by the normalization 
condition: $z(0)=0$. For the value of the first derivative of $f(x)$ 
at the origin we will take the condition $f'(0)=\gamma$, where 
$\gamma$ is a free parameter. The anomalous dimension $\eta(t)$ at a 
fixed point becomes the critical exponent $\eta_{*}$. 
                               
\section{Fixed points and critical exponents}

To solve Eqs. (\ref{eqfp1}), (\ref{eqfp2}) for $d=2$ we consider 
the recursive numerical method already tested for $d=3$ 
\cite{BHLM}. The physical fixed point solutions 
$f_{*}(x)$, $z_{*}(x)$ at the fixed point value $\eta=\eta_{*}$ 
are regular for $x > 0$ and have a certain asymptotic 
behavior as $x\rightarrow +\infty$. Thus 
the natural method for finding the correct numerical 
solution is to select those which can be integrated as 
far as possible in $x$. A generic solution will end at a 
sharp singularity for a finite value of $x$. 
The difficulty lies in the nonlinear and stiff nature of 
the equations and the need to fine tune $\eta$ and $\gamma$. 
This makes the direct integration of the system too hard.
One way out is to solve it recursively.   
 
Unlike the case $d=3$ studied in a number of articles \cite{BHLM,D3}, 
one faces an additional difficulty in two dimensions. 
It is not possible to start the iterative procedure by setting 
in Eq.\ (\ref{eqfp1}) $z=0$ and $\eta=0$ as it is prescribed by 
the consistency of the leading approximation. For $d=2$ the 
Polchinski equation in the leading order has only 
periodic or singular solutions for all values of $\gamma$. 
To overcome this difficulty one has to consider  
$\eta\not=0$ as the initial value to start the iterations. 
Consequently, an analysis of the leading order Polchinski equation 
\begin{equation}
f''-2ff'+{\Delta^+}f+{\Delta^-}xf'=0,      \label{eqfp01}
\end{equation}
with the initial conditions $f(0)=0$, $f'(0)=\gamma$ 
for non-zero $\eta$ is required. 

We studied Eq.\ (\ref{eqfp01}) for $d=2$ numerically for 
a wide range of values of $\eta$ and $\gamma$. Our results 
show that for each $0<\eta\leq 1$ we can fine tune $\gamma$ 
in such a way as to obtain a non-trivial regular  
fixed point solution. The set of such values $(\eta, \gamma)$ 
form a discrete series of continuous lines $\eta(\gamma)$ 
(see Fig. 1). 
In fact simple arguments can be presented which explain the 
appearance of the lines in the parameter space corresponding 
to regular fixed-point solutions $f_{*}(x)$. Let 
$x_{0}(\eta,\gamma)$ denote position of the pole of a generic 
solution of Eq.\ (\ref{eqfp01}). Suppose that for some values 
$(\eta',\gamma')$ the solution is regular, i.e. 
$x_{0}(\eta',\gamma')=+\infty$. Let us take another value 
$\eta''$ sufficiently close to $\eta'$. Assuming that the 
function $x_{0}(\eta,\gamma)$ is continuous, it is clear that 
there should exist the value $\gamma''$ such that again 
$x_{0}(\eta'',\gamma'')=+\infty$. Hence there is a line of 
the "constant value" $x_{0}(\eta,\gamma)=+\infty$ in the parameter 
space. When we move along a fixed line the solutions $f_{*}(x)$ 
do not change their shape significantly. Moreover, their shape 
follows a regular pattern when passing from one curve to the 
other similar to solutions obtained by Morris \cite{TM}.
This can be considered as a sign for the existence of the 
infinite discrete set of fixed points corresponding to the 
minimal unitary series of conformal models. We also would 
like to note that for $-2 < \eta < 6$ and $-1 < \gamma < 1$ 
there are no other non-trivial fixed-point solutions besides 
the ones corresponding to the lines discussed here.  
   
\vspace{-1cm}  
\begin{figure}[ht]
\epsfxsize=1.0\hsize
\epsfbox{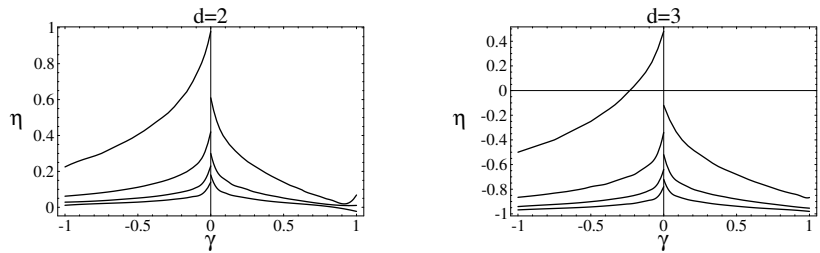}
\vspace{-1.2cm}
\caption{Non-trivial fixed-point lines for $d=2$ and $d=3$. 
For clarity, only the first 7 lines are shown.}
\end{figure}
\vspace{-0.1cm}
We would like to note that a similar picture takes place in 
other dimensions. For $d=3$ we found the same discrete 
set of lines in the $(\gamma, \eta)$-plane corresponding 
to regular solutions of Eq.\ (\ref{eqfp01}), but in this 
case they are situated in the interval $-1<\eta\leq 1/2$ 
(see Fig. 1). As one can see there is a line 
(upper line in Fig. 1) which crosses the $\gamma$-axis at 
$\gamma_{*}=-0.229\ldots$. This is the value of the parameter 
$\gamma$ for which a non-trivial fixed point solution of the 
Polchinski equation was found in the leading order (local 
potential approximation) \cite{BHLM}. The important 
observation is that there is only one line in the   
$(\gamma, \eta)$-plane with positive values of $\eta$. Since 
according to general arguments at physical fixed points 
$\eta_{*}>0$, this suggests that for $d=3$ there is only one 
non-trivial fixed point. This is the Wilson-Fischer fixed point 
found in numerous previous studies \cite{D3}. 

Recall that for $d=2$ all the lines are 
situated in the $\eta > 0$ half-plane, hence one can expect 
an infinite number of non-trivial fixed points. 
One more remark is relevant here. By a 
simple scaling analysis of Eq.\ (\ref{eqfp01}) it can be shown 
that there is a certain mapping between the lines $\eta(\gamma)$ 
in different dimensions. When we pass from one dimension to another 
the line experiences a vertical shift and scaling transformation. 
More details about this mapping will be presented elsewhere.   

We now pass to the study of the system (\ref{eqfp1}), (\ref{eqfp2}). 
We have seen that there are families of solutions of 
Eq.\ (\ref{eqfp01}) corresponding to a given fixed point. It 
turns out that when the second equation of the system is taken 
into account, this degeneracy disappears. We solved the system 
using the following iteration procedure developed by Ball et al.  
\cite{BHLM}. First we set $z(x)=0$, choose some initial 
value $\eta=\eta_{0}$ and fine tune $\gamma$ to the value 
$\gamma={\gamma_0}$ corresponding to the regular solution 
$f_{0}(x)$ of the first equation (\ref{eqfp1}) (or (\ref{eqfp01})). 
Of course, the point $(\eta_{0},\gamma_{0})$ lies on one of the 
lines described above (see Fig. 1)). As the next step we insert the 
function $f_{0}(x)$ into the second equation (\ref{eqfp2}) for 
a fixed $B$ and fine tune $\eta$ to the value $\eta={\eta_1}$ 
for which a regular solution $z_{1}(x)$ exists. Then we substitute 
$z_{1}(x)$ and $\eta_{1}$ into the first equation of the system 
and find a regular solution for a fixed value of $A$ thus 
obtaining a new value $\gamma={\gamma_1}$ and a new function 
$f_{1}(x)$. We repeat this process keeping $A$ and $B$ fixed. 
As a result a sequence of functions $f_0 (x)$, $z_1 (x)$, 
$f_1 (x)$, $z_2 (x), \ldots$, and a sequence of numbers 
$\gamma_0$, $\eta_1$, $\gamma_1$, $\eta_2, \ldots$, 
are obtained, and we test them for convergence.

For $d=3$ and for the relevant line, associated 
with the Wilson-Fischer fixed point, we confirmed the 
results by Ball et al. \cite{BHLM}. The new feature in 
our calculations is that we took $\eta_{0} \neq 0$ as the  
initial value of the iterating procedure, whereas in \cite{BHLM} 
only $\eta_{0}=0$ was considered.
We conclude that the numerical method converges and that 
the rate of convergence is controlled by $A$ for fixed 
$\eta_0$. The best $A$ was shown to correspond to 
the inflexion point where $\Delta\eta=
{\eta_2}-{\eta_1}$ changes sign. The important observation is 
that the final values $\eta_{*}$ and $\gamma_{*}$ to which the 
iterations converge (i.e. the fixed-point values) do not 
depend on the initial value $\eta_{0}$. When $\eta_0$ is closer 
to the fixed-point value the rate of convergence is of course 
faster. The final value $\eta_{*}$ depends on $B$ linearly. 
For $A=0$ the two equations decouple 
and there is no need for iterations to find a solution 
of the system. We just have to adjust $B$ such that 
${\eta_1}={\eta_0}$. For the Wilson-Fischer fixed 
point ${\eta_0}={\eta_1}=0.04$ we found $B=0.666768\ldots$ and 
$\gamma_{*}={\gamma_0}=-0.197435\ldots$.

\vspace{-1cm}
\begin{figure}[ht]
\epsfxsize=1.0\hsize
\epsfbox{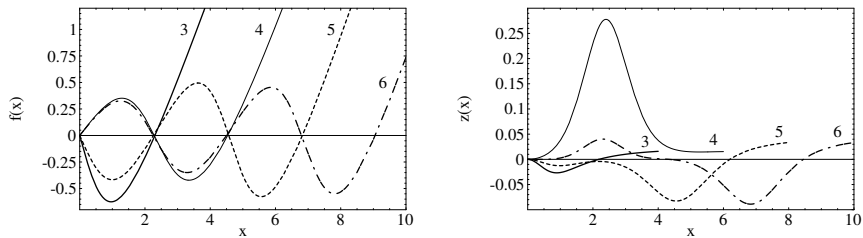} 
\vspace{-1.5cm}
\caption{The solutions $f(x)$ and $z(x)$ for $p=3,4,5,6$.}
\end{figure}
\vspace{-0.1cm}
For $d=2$ the situation is totally different. 
For any line $\eta (\gamma)$ we start with and 
the initial value $(\eta_0, \gamma_{0})$ 
the iterative procedure turns out to be divergent if 
$A \not=0$. Only for $A=0$ we have been able to find a 
solution to the now decoupled system (\ref{eqfp1}), (\ref{eqfp2}) 
by adjusting $B$. Similar to the case $d=3$ we have not found any  
natural criteria to select the value of $B$ since for each line $B$ 
depends monotonically on $\eta$ decreasing as $\eta\rightarrow 0$.
To determine the fixed-point solutions, corresponding to the 
minimal unitary series of conformal field theories, we have fixed 
the value of $B$ by a fit to the series of exact values for 
the anomalous scaling dimension $\eta_{*}$. In this way we found 
$B \approx 0.25$. The fixed-point solutions for $f_{*}(x)$ display    
regular behaviour and are reminiscent of those obtained by Morris 
\cite{TM}. In particular they have $p-2$ extrema,   
$p=3,4,\ldots$ (see Fig. 2). The fixed 
point solutions for $z_{*}(x)$ have the same pattern of extrema, 
though their profiles are different from those of Morris. 

Next for $B=1/4$ and and corresponding values $\eta=\eta_{*}$ we 
have calculated the critical exponent $\nu$. For this we considered  
perturbations of the functions $f(x)$ and $z(x)$ around the 
fixed-point solutions,
\[
f(x)={f_*}(x)+{\sum_{n=1}^\infty}{g_n}(x){e^{{\lambda_n}t}},
\quad
z(x)={z_*}(x)+{\sum_{n=1}^\infty}{h_n}(x){e^{{\lambda_n}t}}, 
\]
and substituted them into Eqs.\ (\ref{eq1}), (\ref{eq2}). After 
linearization we obtained the system
\vspace{-0.3cm} 
\bea
{g_n}''-2{g_n}{f_*}'-2{f_*}{g_n}'+{\Delta_{*}^{+}}{g_n}+
{\Delta_{*}^{-}}x{g_n}'&=& {\lambda_n}{g_n}, \nonumber \\
{h_n}''-2{f_*}{h_n}'-2{g_n}{z_*}'-4{f_*}'{h_n}-4{g_n}'{z_*}
\qquad&& \nonumber \\
+{\Delta_{*}^{-}}x{h_n}'-{\eta_*}{h_n}+ 2B{f_*}'{g_n}'&=&{\lambda_n}{h_n},
\nonumber
\eea
where $\Delta_{*}^{+}$ and $\Delta_{*}^{-}$ are calculated for 
$\eta=\eta_{*}$. 

For $Z_2$ perturbations the initial conditions are 
${g_n}(0)=0$, ${h_n}'(0)=0$. We also imposed the 
normalization condition ${g_n}'(0)=1$. Away from the 
fixed point (but sufficiently close to it) we relaxed ${h_n}$ 
to be different from zero, ${h_n}=\delta$. 
Then $\lambda_n$ and $\delta$ were fine tuned so that 
polynomially growing eigenfunctions were obtained.
For $d=2$, $A=0$ and $B=1/4$ we have calculated the critical exponent 
$\nu=1/{\lambda_1}$. The results for $\eta_{*}$ and $\nu$ 
are given in Table 1. We conclude that the Polchinski ERG 
equation gives the values for the critical exponent which 
match quite well the exact conformal field theory 
values \cite{AZ} and the results by Morris \cite{TM}, also 
included in the table. 

We note that the best fit value for $B$ that we have found 
actually corresponds to a well defined subset of the regulating 
functions. An explicit example is $K(x)=(1+ax+b{x^2}){e^{-x}}$,
where $a+2b+1=0$ to ensure ${I_1}=0$ and ${I_0}={\Omega_d}$,  
and $a=-5\pm 2\sqrt{6}$ to give $B=1/4$.

\vspace{0.5cm}
\leftline{
\begin{tabular*}{11cm}{|r|@{\extracolsep{\fill}}c|@{\extracolsep{\fill}}c|
@{\extracolsep{\fill}}c|@{\extracolsep{\fill}}c|@{\extracolsep{\fill}}c|
@{\extracolsep{\fill}}c|@{\extracolsep{\fill}}c|@{\extracolsep{\fill}}c|
@{\extracolsep{\fill}}c|@{\extracolsep{\fill}}c@{\extracolsep{\fill}}|}
\hline
$p$ & 3 & 4 & 5 & 6 & 7 & 8 & 9 & 10 & 11 & 12 \\
\hline
$\eta_{*}$ & .204 & .183 & .0865 & .0708 & .0496 & .0413 &.0328 
& .0281 & .0234 & .0206 \\
\hline
$\eta_{\hbox{\tiny{CFT}}}$ & .25 & .15 & .10 & .0714 & .0536 & .0417 
& .0333 & .0273 & .0227 & ..0192 \\
\hline
$\eta'$ & .309 & .200 & .131 & .0920 & .0679 & 
.0521 & .0412 & .0334 & 0277 & .0233 \\
\hline
$\nu$ & 1.41 & .560 & .527 & .521 & .515 & .512 & .509 & .508 & .507 
& .506 \\
\hline
$\nu_{\hbox{\tiny{CFT}}}$ & 1 & .556 & .536 & .525 & .519 & .514 
& .511 & .509 & .508 & .506 \\
\hline
$\nu'$ & .863 & .566 & .545 & .531 & .523 & .517 & 
.514 & .511 & .509 & .508 \\
\hline
\end{tabular*}
}

\vspace{0.5cm}

\leftline{Table 1: The critical exponents $\eta_{*}$ and $\nu$ 
obtained, as compared to the }

\hspace{0.65cm} values obtained by Morris \cite{TM} ($\eta'$, $\nu'$) 
and the exact results of 
 
\hspace{0.65cm} conformal field theory ($\eta_{\hbox{\tiny{CFT}}}$, $\nu_{\hbox{\tiny{CFT}}}$).
  
\section{Discussion and conclusions}

In this work we have studied the solutions of the 
Polchinski ERG equation for an effective $Z_2$-symmetric scalar 
field theory in the two-dimensional space $R^{2}$. 
We have seen that this equation provides a reliable 
non-perturbative evidence for the existence of the fixed-point 
solutions corresponding to the minimal unitary series of 
conformal field theories and allows to calculate the anomalous 
dimension and the critical exponents with good accuracy. 
This constitutes another positive test of the power
of the ERG approach. At the same time our studies are complementary 
to similar calculations within the ERG approach based on 
the equation for the Legendre action \cite{TM}.

As mentioned above, in the leading order of the derivative expansion 
(local potential approximation) the consistent value for $\eta$ is 0 
and only periodic sine-Gordon type fixed-point solutions can be 
obtained. However, we have found that there are continuous 
families of fixed-point solutions corresponding to a series of 
lines $\eta(\gamma)$ in the $(\gamma, \eta)$ plane which have a   
part with positive values $\eta>0$. It was also argued 
that these lines correspond to the multicritical fixed points 
of the theory. It is by taking into account the second order 
in the derivative expansion that we find isolated fixed-point solutions. 
We have found the first 10 points out of the infinite series and  
calculated the critical exponent  $\nu$ for them. 
The results depend on the choice of the regulating function. 
The value of $B$, for which a regular solution exists, depends linearly 
on $\eta$, so the criterium of minimal sensitivity cannot be applied to 
fix $B$. The best fit to the conformal field 
theory values of $\eta$ gives us $A=0$, $B \approx 0.25$. No other 
regulating functions have been found to work. For $d=2$ whenever 
$A\not=0$, the iterative procedure is not seen to converge.
In fact, for $A=0$ there is no need for iterations since the 
two equations of the Polchinski approach decouple. 
This is in sharp contrast to the case of $d=3$ where convergence 
was checked for $A\not=0$ \cite{BHLM}. 
For the values $A=0$, $B=1/4$, giving the best fit, our results 
are comparable in accuracy and sometimes 
better (the accuracy also increases with multicriticality) than 
those of Morris \cite{TM}. We would like to note that fixing 
$B$ by the best fit to exact results for the anomalous dimensions 
$\eta$ is reminiscent of fixing the renormalization scheme 
dependence in the perturbative renormalization group. It is also 
similar to fixing the regulator by the condition of the 
reparametrization 
invariance for the ERG equation for the Legendre action 
(that corresponds to the limiting case of $A=0$, $B = 
\infty$) \cite{TM}. 

Another important point we would like to mention is 
that in our analysis we have not found  
non-trivial fixed points other than those corresponding 
to the minimal models. This is what one expects from 
Zamolodchikov's $c$-theorem~\cite{AZ1}. The conclusion 
is already clear from the analysis of the leading order Polchinski 
equation for $\eta\not=0$ when the lines in the 
$(\gamma,\eta)$-plane corresponding to non-trivial fixed points 
are plotted. 

As final comments we would like to mention that to obtain an 
estimate of the error of our numerical results in Table 1 
one needs to carry out the calculations to the next order 
of the derivative expansion.
It would be also interesting to expand the analysis 
for higher dimension operators. 

\vspace{0.3cm}

\leftline{\bf Acknowledgements}

\vspace{0.15cm}

We would like to thank Tim Morris and Jos\'e Latorre 
for some fruitful discussions during the Workshop. 
We also acknowledge financial support by Funda\c {c}\~ao para a 
Ci\^encia e a Tecnologia under grant number CERN/S/FAE/ \\
1177/97.
Yu.K.\ acknowledges financial support from 
fellowship PRAXIS XXI \\
/BCC/4802/95.
R.N.\ acknowledges financial support from 
fellowship PRAXIS XXI/BPD/14137/97.


\vspace{0.5cm}

\leftline{\bf References}

\begin{thebibliography}{15}

\bibitem[1]{KW} 
K. Wilson, Rev. Mod. Phys. 47 (1975) 773; \\
K. Wilson and J. Kogut, Phys. Rep. 12 (1974) 75.

\bibitem[2]{SW} 
S. Weinberg, in Understanding the Fundamental Constituents of Matter, 
Erice 1976, edited by A. Zichichi (Plenum, 1978).

\bibitem[3]{WH} 
F.J. Wegner and A. Houghton, Phys. Rev. A8 (1973) 401.

\bibitem[4]{JP} 
J. Polchinski, Nucl. Phys. B231 (1984) 269.

\bibitem[5]{YK} 
Y. Kubyshin, Int. J. Mod. Phys. B12 (1998) 1321.

\bibitem[6]{TM1} 
T.R. Morris, Int. J. Mod. Phys. B12 (1998) 1343; In: "Zakopane 1997. 
New developments in quantum field theory" (NATO Workshop on Theoretical 
Physics, Zakopane, 1997), p. 147 (hep-th/9709100); preprint, hep-th/9802039.

\bibitem[7]{AZ} 
A.B. Zamolodchikov, Yad. Fiz. 44 (1986) 821; 
J.L. Cardy, in Phase Transitions and Critical Phenomena, 
Eds. C. Domb and J.L. Lebowitz (Academic Press, 1987) Vol 11.

\bibitem[8]{BPZ} 
A.A. Belavin, A.M. Polyakov and A.B. Zamolodchikov, Nucl. Phys. B241 
(1984) 333; 
D. Friedan, Z. Qiu and S. Shenker, Phys. Rev. Lett. 52 (1984) 1575.

\bibitem[9]{TM} 
T.R. Morris, Phys. Lett. B345 (1995) 139.

\bibitem[10]{TM2}
T.R. Morris, Int. J. Mod. Phys. A9 (1994) 2411. 

\bibitem[11]{BHLM} 
R.D. Ball, P..E. Haagensen, J.I. Latorre and E. Moreno, 
Phys. Lett. B347 (1995) 80.

\bibitem[12]{BT} 
R.D. Ball and R.S. Thorne, Ann. Phys. 236 (1994) 117.

\bibitem[13]{D3}
A. Hasenfratz and P. Hasenfratz, Nucl. Phys. B270 (1986) 687; 
T.R. Morris, Phys. Lett. B329 (1994) 241; 
P.E. Haagensen, Yu. Kubyshin, J.I. Latorre and E. Moreno, 
Phys. Lett. B323 (1994) 330; 
N. Tetradis and C. Wetterich, Nucl. Phys. B422 (1994) 541; 
J. Comellas, Nucl. Phys. B509 (1998) 662. 

\bibitem[14]{AZ1} 
A.B. Zamolodchikov, JETP Lett. 43 (1986) 730. 

\end{thebibliography}
\end{document}